\global\def\draftcontrol{0}

   \def\versionno{ n=2 bulk viscosity }

\catcode`\@=11

\expandafter\ifx\csname draftcontrol\endcsname\relax\global\def\draftcontrol{0}
\fi

{\count255=\time\divide\count255 by 60
\xdef\hourmin{\number\count255}
\multiply\count255 by-60\advance\count255 by\time
\xdef\hourmin{\hourmin:\ifnum\count255<10 0\fi\the\count255}}
\def\draftdate{\number\month/\number\day/\number\year\ \ \ \hourmin }

\newcommand\makepapertitle{\par
  \begingroup
    \renewcommand\thefootnote{\@fnsymbol\c@footnote}%
    \def\@makefnmark{\rlap{\@textsuperscript{\normalfont\@thefnmark}}}%
    \long\def\@makefntext##1{\parindent 1em\noindent
            \hb@xt@1.8em{%
                \hss\@textsuperscript{\normalfont\@thefnmark}}##1}%
     \newpage
     \global\@topnum\z@   
     \@makepapertitle
     \thispagestyle{empty}\@thanks
  \endgroup
  \setcounter{footnote}{0}%
  \global\let\thanks\relax
  \global\let\makepapertitle\relax
  \global\let\@makepapertitle\relax
  \global\let\@thanks\@empty
  \global\let\@author\@empty
  \global\let\@date\@empty
  \global\let\@title\@empty
  \global\let\title\relax
  \global\let\author\relax
  \global\let\date\relax
  \global\let\and\relax
  \def\version{\let\version\@version\@gobble}
}
\def\@makepapertitle{%
  \newpage
   \ifnum\draftcontrol=1 {}
   \version\versionno
   \vskip 3em%
   \else
   \hfill\hbox to 3cm {\parbox{4cm}{\@pubnum}\hss}%
   \vskip 3em%
   \fi
   \begin{center}%
   \let \footnote \thanks
     {\LARGE {\@title}}%
     \vskip 1.5em%
     {\normalsize
       \lineskip .5em%
       \begin{tabular}[t]{c}%
         \@author
       \end{tabular}\par}%
     \vskip 1.5em%
     {\@bstract}%
     \end{center}%
     \vskip 1.5em
     \@date%
   \par
}

\gdef\@pubnum{}
\def\pubnum#1{%
  \gdef\@pubnum{#1}}

\gdef\@bstract{}
\def\Abstract#1{%
  \gdef\@bstract{%
   \parbox{\textwidth-0pc}{%
   \centerline{\bf Abstract}\penalty1000%
\kern.2cm%
\noindent
\renewcommand\baselinestretch{1.0}%
{#1}}}
}

\def\ps@paper{\let\@mkboth\@gobbletwo%
     \ifnum\draftcontrol=1
    \def\@oddfoot{\hbox to \textwidth{\tiny \versionno \hfil\tiny\draftdate}%
    \hskip -\textwidth \hbox to \textwidth{\hfil\rm\thepage\hfil}}%
     \else\def\@oddfoot{\hbox to \textwidth{\hfil\rm\thepage\hfil}}
     \fi
     \let\@evenfoot\@oddfoot
}

\def\body{\clearpage
          \pagestyle{paper}
    }

\def\@version#1{\ifnum\draftcontrol=1
\typeout{}\typeout{#1}\typeout{}
\vskip3mm\centerline{\hbox{\fbox{\normalsize{\tt DRAFT -- #1 -- }
                   {\draftdate}}}}\vskip3mm
\fi}
\let\version\@version
\long\def\eqlabel#1{\ifnum\draftcontrol=1
                    \tag@false  
                    \tag*{(\theequation) \hbox to -0.2cm{\hspace{0cm}\small{#1}\hss}}
                    \refstepcounter{equation}
                    \edef\@currentlabel{\theequation}
                    \ltx@label{#1}          
                    \else
                    \label{#1}
                    \fi
                    }
\let\st@bibitem\@bibitem
\let\st@lbibitem\@lbibitem
\ifnum\draftcontrol=1
  \def\@bibitem#1{%
    \st@bibitem{#1}\a@@label{#1}\ignorespaces}
  \def\@lbibitem[#1]#2{%
    \st@lbibitem[#1]{#2}\a@@label{#2}\ignorespaces}
  \def\a@@label#1{%
    \gdef\a@lab{\smash{\normalfont\small#1}}
    \ifvmode
      \if@inlabel
        \global\setbox\@labels\hbox{%
          \llap{\a@lab\let\a@lab\relax
                \kern\@totalleftmargin\kern\marginparsep}%
          \box\@labels}%
      \fi
    \fi}
\fi

\documentclass[12pt,letterpaper]{article}

\usepackage{amsmath,amssymb,array,calc,epsfig}
\usepackage{graphicx}
\usepackage{psfrag,verbatim,bm}

\ifnum\draftcontrol=1
\fi

\renewcommand\baselinestretch{1.25}
\renewcommand\section{\@startsection {section}{1}{\z@}%
                                   {-3.5ex \@plus -1ex \@minus -.2ex}%
                                   {2.3ex \@plus.2ex}%
                                   {\normalfont\large\bfseries}}
\renewcommand\subsection{\@startsection{subsection}{2}{\z@}%
                                   {-3.25ex\@plus -1ex \@minus -.2ex}%
                                   {1.5ex \@plus .2ex}%
                                   {\normalfont\normalsize\bfseries}}
\renewcommand\subsubsection{\@startsection{subsubsection}{3}{\z@}%
                                   {-3.25ex\@plus -1ex \@minus -.2ex}%
                                   {1.5ex \@plus .2ex}%
                                   {\normalfont\normalsize\it}}
\renewcommand\paragraph{\@startsection{paragraph}{4}{\z@}%
                                   {-3.25ex\@plus -1ex \@minus -.2ex}%
                                   {1.5ex \@plus .2ex}%
                                   {\normalfont\normalsize\bf}}

\numberwithin{equation}{section}

\def\ie{{\it i.e.}}

\def\revise#1       {\raisebox{-0em}{\rule{3pt}{1em}}%
                     \marginpar{\raisebox{.5em}{\vrule width3pt\
                     \vrule width0pt height 0pt depth0.5em
                     \hbox to 0cm{\hspace{0cm}{%
                     \parbox[t]{4em}{\raggedright\footnotesize{#1}}}\hss}}}}

\newcommand\nxt[1]  {\\\fnxt#1}

\def\calc         {{\cal C}}

\def\cale         {{\cal E}}
\def\calf         {{\cal F}}

\def\call         {{\cal L}}
\def\calm         {{\cal M}}
\def\caln         {{\cal N}}
\def\calo         {{\cal O}}
\def\calp         {{\cal P}}

\def\cals         {{\cal S}}

\def\del          {\partial}

\def\sqr#1#2{{\vcenter{\vbox{\hrule height.#2pt
 \hbox{\vrule width.#2pt height#1pt \kern#1pt
 \vrule width.#2pt}\hrule height.#2pt}}}}

\newcommand{\ft}[2]{{\textstyle{\frac{#1}{#2}}}}

\def\a{\alpha}
\def\b{\beta}
\def\l{\lambda}
\def\w{\omega}
\def\dd{\delta}
\def\r{\rho}
\def\c{\chi}
\newcommand{\qq}{\mathfrak{q}}
\newcommand{\ww}{\mathfrak{w}}

\catcode`\@=12

\begin{document}


\title{\bf Bulk viscosity of $\caln=2^*$ plasma}
\pubnum
{UWO-TH-08/18
}

\date{December 2008}

\author{
Alex Buchel$ ^{1,2}$ and Chris Pagnutti$ ^{1}$\\[0.4cm]
\it $ ^1$Department of Applied Mathematics\\
\it University of Western Ontario\\
\it London, Ontario N6A 5B7, Canada\\
\it $ ^2$Perimeter Institute for Theoretical Physics\\
\it Waterloo, Ontario N2J 2W9, Canada\\
}

\Abstract{
We use gauge theory/string theory correspondence to study the bulk
viscosity of strongly coupled, mass deformed $SU(N_c)$ $\caln=4$
supersymmetric Yang-Mills plasma, also known as $\caln=2^*$ gauge
theory. For a wide range of masses we confirm the bulk viscosity bound
proposed in \cite{b1}. For a certain choice of masses, the theory
undergoes a phase transition with divergent specific heat $c_V\sim
|1-T_c/T|^{-1/2}$. We show that, although bulk viscosity rapidly grows
as $T\to T_c$, it remains finite in the vicinity of the critical
point.
}

\makepapertitle

\body

\version\versionno

\section{Introduction}
In \cite{m9711} Maldacena proposed that $SU(N_c)$ $\caln=4$ superconformal Yang-Mills (SYM) theory on $R^{3,1}$
is dual to type IIB string theory on (a Poincare patch of) $AdS_5\times S^5$ with $N_c$ units of the Ramond-Ramond
five-form flux through the $S^5$. Assuming that such a duality holds exactly at a superconformal fixed point, 
it must also hold for {\it any} relevant (in the infrared) deformation of a fixed point. Specifically, we might 
generate non-conformal examples of gauge theory/string theory correspondence by simply mapping  
the mass deformation on the SYM side to the string theory side. For infinitesimal  supersymmetric mass deformations this was done in 
\cite{ps} using the operator/state correspondence in AdS/CFT \cite{gkp,w1}. Extending an infinitesimal mass deformation
to a finite one proved to be extremely difficult. In fact, only one particular deformation was shown to be integrable
in the 't Hooft limit $N_c\to\infty$, $g_{YM}\to 0$ (with $\l\equiv N_c g_{YM}^2$ kept constant), and for large 't Hooft coupling 
$\l\gg 1$ \cite{pw} (PW). In \cite{bpp} it was shown\footnote{Related discussion appeared in \cite{ejp}.} that PW massive 
deformation is dual to giving the same mass to two chiral multiplets of the parent $\caln=4$ SYM at a specific point on its Coulomb
branch\footnote{Extending the mass deformation duality to all Coulomb branch of the $\caln=4$ SYM is a difficult open problem.}. 
Such a mass deformation breaks supersymmetry down to $\caln=2$. The resulting gauge theory, commonly referred to as 
$\caln=2^*$ gauge theory, can be solved nonperturbatively \cite{dw}. The agreement \cite{bpp} between the dual gravitational description 
of the massive $\caln=2$ gauge theory \cite{pw} and its exact field-theoretic solution \cite{dw}  provides a highly nontrivial 
check of gauge theory/string theory correspondence in a non-conformal setting.

Once the duality is established for a supersymmetric ground state, following \cite{wthermal} it can be extended to correspondence 
involving a thermal equilibrium state of the gauge theory: a thermally equilibrium state of a gauge theory is dual to a Schwarzschild
black brane solution in the supergravity description, with the temperature $T$ given by the Hawking temperature of the 
black brane.  In the context of $\caln=2^*$ theory there is an interesting subtlety: 
the supersymmetric mass deformation involves deformation of the parent SYM by operators of different canonical dimensions, a dimension-two 
operator for the bosonic components of massive chiral multiplets and a dimension-three operator for the fermionic components 
of massive chiral multiplets. Such operators are mapped to {\it different} scalar gravitational modes of the effective 
five-dimensional gravitational description \cite{pw}. The coefficients of the non-normalizable modes of these scalars encode  
the bosonic 
$m_b$ and fermionic $m_f$ masses respectively \cite{ps}. Although the vacuum state supersymmetry requires $m_b=m_f$, a thermal 
state breaks the supersymmetry anyway; thus, we can study a phase diagram of $\caln=2^*$ gauge theory with $m_b\ne m_f$ \cite{bl}.
A full ten-dimensional type IIB supergravity solution dual to $\caln=2^*$ gauge theory for generic $\left(T,m_b,m_f\right)$
was constructed in \cite{bl}. It was shown there (see also \cite{bc}) 
that any such state (including small fluctuations about it) can be described 
within effective five-dimensional gauged supergravity presented in \cite{pw}. 

Supergravity equations of motion derived in \cite{bl} describing a thermal state of $\caln=2^*$ 
gauge theory
are too difficult to solve analytically.
If fact, in \cite{bl} these equations were solved only in high-temperature limit $m_b/T\ll 1$ and $m_f/T\ll 1$. Additional analysis 
of the $\caln=2^*$ phase diagram required numerical work. For two special cases, \ie, $(susy)\equiv \{m_b=m_f\}$ and 
$(bosonic)\equiv \{m_f=0\,, m_b\ne 0\}$, 
such numerical analysis were performed in \cite{bdkl}. Using the holographic renormalization of the theory \cite{bhyd}, 
the energy density $\cale$, the free energy density $\calf$ and the 
entropy density $s$ of $\caln=2^*$ strongly coupled plasma was computed. It was shown in \cite{bdkl} that  the basic 
thermodynamic relation $\calf=\cale-T s$ is satisfied exactly, while the first law of thermodynamics 
$d\cale=T ds$ is exactly satisfied for the high temperature analytic solution of \cite{bl}, and is satisfied numerically for 
$(susy)$ and $(bosonic)$ thermal states with an  accuracy of $\sim 0.1\%$ and  $\sim 0.01\%$ correspondingly.
The latter provides a highly nontrivial check on  our identification of the bosonic and fermionic masses in 
dual supergravity (see \cite{bdkl} for details). 

An interesting critical point was found in the numerical analysis of the $(bosonic)$ thermal state of the $\caln=2^*$ plasma 
\cite{bdkl}: for $T< T_c\approx m_b/2.29(9)$ $\caln=2^*$ plasma becomes unstable with respect to energy density fluctuations.
Specifically, precisely at $T=T_c$ the speed of sound waves squared $c_s^2$ vanishes. 
A perturbative instability of this type is a defining feature of a second order phase transition. Since $c_s^2\propto (T-T_c)^{1/2}$,
the specific heat $c_V$ diverges as $|1-T_c/T|^{-1/2}$, suggesting that  
such a critical point is in the universality class of the  mean-field tricritical point. Physically, the existence of perturbative 
instability in $\caln=2^*$ plasma at low temperatures is not surprising. Indeed, once $m_b\ne m_f$ the supersymmetry is broken, 
and the theory is guaranteed to be stable only at high temperatures. It was conjectured in \cite{bdkl} that $\caln=2^*$ plasma 
would have a critical point 
\begin{equation}
T_c=T_c\left(\nu\equiv \frac{m_f^2}{m_b^2}\right)\,,
\eqlabel{defnu}
\end{equation} 
as long as $\nu<1$.

Shear viscosity of $\caln=2^*$ plasma was computed in \cite{u1,bhyd}; it was shown to satisfy the universal bound \cite{u1,u2,u3}
\begin{equation}
\frac \eta s=\frac{1}{4\pi}\,.
\eqlabel{etas}
\end{equation} 
Transport properties of $\caln=2^*$ plasma were further studied in \cite{bbs}:
\nxt equations of motion describing sound quasinormal modes of $\caln=2^*$ black brane with 
arbitrary momentum $\qq\equiv |\vec q|/(2\pi T)$
and frequency $\ww\equiv \w/(2\pi T)$ were obtained;
\nxt these equations were solved analytically in the hydrodynamic limit $\qq\ll 1\,, \ww\ll 1$, and for high temperatures 
$T\gg \{m_b,m_f\}$;  
\nxt following \cite{ks}, the dispersion relation for the lowest quasinormal mode of the $\caln=2^*$ 
black brane was identified with the dispersion relation of the sound waves in strongly coupled 
$\caln=2^*$ plasma:
\begin{equation}
\ww=c_s\ \qq-2\pi i\ \frac \eta s \left(\frac 23+\frac{\zeta}{2\eta}\right)\ \qq^2+\calo\left(\qq^3\right) \,,
\eqlabel{sound}
\end{equation}
where $\zeta$ is the plasma bulk viscosity;
\nxt to leading order in $\{m_f/T, m_b/T\}$ it was found that 
\begin{equation}
c_s=\frac{1}{\sqrt{3}}\left(1-\frac{\left[\Gamma\left(\frac 34\right)\right]^4}{3\pi^4}\ 
\left(\frac{m_f}{T}\right)^2-\frac{1}{18\pi^4}\left(\frac{m_b}{T}\right)^4+\cdots\right)\,,
\eqlabel{cs} 
\end{equation}
in precise agreement with the speed of sound obtained from the analytic equation of state of $\caln=2^*$ plasma at high 
temperature\footnote{This provides a highly nontrivial consistency check on our analysis of the quasinormal modes.} 
\begin{equation}
c_s^2=-\frac{\del \calf}{\del \cale}\,;
\eqlabel{eos}
\end{equation}
\nxt using \eqref{etas}, \eqref{cs}, to leading order in $\left(\frac 13 -c_s^2\right)$ it was found that 
\begin{equation}
\frac{\zeta}{\eta}\bigg|_{m_f=0}=\frac{\pi^2\beta_b^\Gamma}{16}\left(\frac 13-c_s^2\right)+{\cal O}\left(\left[\frac 13-c_s^2\right]^2\right)\,,
\eqlabel{n2r1}
\end{equation}
where $\beta_b^\Gamma\approx 8.001$; 
\begin{equation}
\frac{\zeta}{\eta}\bigg|_{m_b=0}=\frac{3\pi\beta_f^\Gamma}{2}\left(\frac 13-c_s^2\right)+{\cal O}\left(\left[\frac 13-c_s^2\right]^2\right)\,,
\eqlabel{n2r2}
\end{equation}
where $\beta_f^\Gamma\approx 0.66666$ \cite{mistake}.

In this paper we extend analysis of \cite{bbs,b1} to general mass deformations, \ie, apart from cases of 
$(bosonic)\Leftrightarrow \nu=0$
and $(susy)\Leftrightarrow \nu=1$ thermal states of $\caln=2^*$ plasma, and for wide range of temperatures. Our goal is twofold:
\nxt first, we would like to test the bulk viscosity bound conjecture of \cite{b1} in more general setting:
\begin{equation}
\frac{\zeta}{\eta}\ge 2 \left(\frac 13-c_s^2\right)\,;
\eqlabel{bound}
\end{equation}
\nxt second, we  would like to compute 
\begin{equation}
\frac{\zeta}{\eta}\bigg|_{critical}\equiv \frac{\zeta}{\eta}\bigg|_{T=T_c(\nu)}\,,
\eqlabel{zetac}
\end{equation}
for $0<\nu<1$.

The paper is organized as follows. In the next section we outline the equations of motion and the boundary conditions 
for the $\caln=2^*$ black brane background and its hydrodynamic quasinormal mode.
 The results of our extensive numerical analysis 
are presented in section 3. We conclude in section 4.

Most technical details are omitted due to their complexity. All equations of motion, their analytic asymptotic solutions,
as well as raw numerical data is available from the authors upon request. We use numerical techniques developed in \cite{abk}.

\section{$\caln=2^*$ black brane and its hydrodynamic quasinormal mode}

We closely follow \cite{bdkl} in discussion of $\caln=2^*$ black brane background, and we (mostly) 
follow notations of \cite{bbs} in discussion of its quasinormal modes.

\subsection{Effective action}
The effective action of the five-dimensional gauged supergravity describing $\caln=2^*$ black brane thermodynamics/hydrodynamics 
is given by
\begin{equation}
\begin{split}
S=&\,
\int_{\calm_5} d\xi^5 \sqrt{-g}\ \call_5\\
=&\frac{1}{4\pi G_5}\,
\int_{\calm_5} d\xi^5 \sqrt{-g}\left[\ft14 R-3 (\del\a)^2-(\del\chi)^2-
\calp\right]\,,
\end{split}
\eqlabel{action5}
\end{equation}
where the potential%
\footnote{We set the five-dimensional gauged
supergravity coupling to one. This corresponds to setting the
radius $L$ of the five-dimensional sphere in the undeformed metric
to $2$.}
\begin{equation}
\calp=\frac{1}{16}\left[\frac 13 \left(\frac{\del W}{\del
\a}\right)^2+ \left(\frac{\del W}{\del \chi}\right)^2\right]-\frac
13 W^2\,
 \eqlabel{pp}
\end{equation}
is a function of $\alpha$ and $\chi$, and is determined by the
superpotential
\begin{equation}
W=- e^{-2\alpha} - \frac{1}{2} e^{4\alpha} \cosh(2\chi)\,.
\eqlabel{supp}
\end{equation}
In our conventions, the five-dimensional Newton's constant is
\begin{equation}
G_5\equiv \frac{G_{10}}{2^5\ {\rm vol}_{S^5}}=\frac{4\pi}{N_c^2}\,.
\eqlabel{g5}
\end{equation}

\subsection{$\caln=2^*$ black brane background}

We parameterize the background metric of the $\caln=2^*$ black brane as 
\begin{equation}
ds_5^2=c_2(x)^2\ \left(-(1-x)^2 dt^2+dx_1^2+dx_2^2+dx_3^2\right)+g_{xx}(x)\ dx^2\,,
\eqlabel{metricx}
\end{equation}
where the radial coordinate $x\in [0,1]$, with $x\to 0_+$ corresponding to the $AdS_5$  boundary
and  $x\to 1_-$ corresponding to a regular Schwarzschild horizon of the black brane.
The metric \eqref{metricx} is supported by  nontrivial profiles of two scalar fields:
\begin{equation}
\rho_6(x)\equiv e^{6\a}\,,\qquad c(x)\equiv \cosh(2\chi)\,.
\eqlabel{scalars}
\end{equation}
Notice that we redefined the gauged supergravity scalars as in \eqref{action5} ---
this is done in order to speed up numerical integration.
We further  introduce 
\begin{equation}
c_2\equiv e^A\,,\qquad A\equiv \ln \hat{\dd}_3-\frac 14 \ln(2x-x^2)+a(x)\,,
\eqlabel{defa}
\end{equation}  
where $\hat{\dd}_3$ is related to the Hawking temperature of the black brane as follows \cite{bdkl}
\begin{equation}
T=\frac{\hat{\dd}_3}{2\pi} \lim_{x\to1_-}e^{-3 a(x)}\,.
\eqlabel{t}
\end{equation}

The asymptotic solution of $\{\r_6,c,a\}$ near the boundary, $x\to 0_+$, takes the form \cite{bdkl}
\begin{equation}
\begin{split}
\r_6=&1+x^{1/2}\left(6\r_{10}+6\r_{11}\ln x\right)+\calo\left(x \ln^2 x\right)\,,\\
c=&1+12 \nu\ x^{1/2}\r_{11}+24\nu \r_{11}\ x\left(\c_{10}+\nu\r_{11}+2\nu\r_{11}\ln x\right)+\calo\left(x^{3/2}\ln^2 x\right)\,,\\
a=&-\frac 23 \nu\r_{11} x^{1/2}+\calo\left(x \ln^2 x\right)\,,
\end{split}
\eqlabel{asbound}
\end{equation}
and near the horizon, $y\equiv 1-x\to 0_+$ \cite{bdkl}
\begin{equation}
\begin{split}
\r_6=\r_0^6+\calo(y^2)\,,\qquad c=\frac{\c_0^4+1}{2\c_0^2}+\calo(y^2)\,,\qquad a=a_0+a_1 y^2+\calo(y^4)\,.
\end{split}
\eqlabel{ashor}
\end{equation}
In \eqref{asbound},\eqref{ashor} we indicated explicitly only terms necessary to unambiguously determine the asymptotic  
black brane geometry for a 
fixed set $\{\mu,\nu\}$. $\r_{11}$ is related to $\mu\equiv \frac{m_b}{T}$ by \cite{bdkl}
\begin{equation}
\r_{11}=\frac{\sqrt{2}}{24\pi^2}e^{-6a_0}\ \mu^2\,.
\eqlabel{r11}
\end{equation}
Notice that for a fixed set $\{\mu,\nu\}$,  $\caln=2^*$ black brane geometry is specified by six parameters:
\begin{equation}
 \{\mu,\nu\}\qquad \Longrightarrow\qquad \biggl\{\r_{10},\c_{10},\r_0,\c_0,a_0,a_1\biggr\}\,,
\eqlabel{parback}
\end{equation}
which is precisely the number of parameters needed to uniquely determine the solution of three second order 
equations of motion for $\{\r_6,c,a\}$. These parameters are functions of $\{\mu,\nu\}$.

We use numerical techniques developed in \cite{abk} to generate data sets 
\begin{equation}
\cals_{background}\equiv\biggl\{\nu;\ \mu;\ \r_{10},\c_{10},\r_0,\c_0,a_0,a_1\biggr\}\,,
\eqlabel{defsb}
\end{equation}
which can be used to study the thermodynamics of $\caln=2^*$ black branes as detailed in \cite{bdkl}. 
The data sets $\cals_{background}$ are available from the authors upon request.
We obtain the following results.
\nxt
For $\nu>1$, in agreement with the conjecture in \cite{bdkl}, we did not find a critical point in  $\caln=2^*$ plasma
down to $\mu\sim 10$, which corresponds to temperatures of order $T\sim m_b/10$. As for $\nu=1$ in \cite{bdkl},
the low temperature ($\mu \sim 10 $) thermodynamics of $\caln=2^*$ black brane can be well approximated by the following 
equation of state 
\begin{equation}
\calf\propto -N_c^2\ T^4\ e^{-\frac{m_{eff}}{T}} \,,
\eqlabel{fnu}
\end{equation}  
where $m_{eff}=m_{eff}(\nu)\sim m_b/10$.
\nxt For $0< \nu<1$, $\caln=2^*$ brane thermodynamics has a critical point $T_c=T_c(\nu)$, such that
\begin{equation}
\frac {s}{c_V}= c_s^2=-\frac{\del \calf}{\del\cale} \propto \pm (T-T_c)^{1/2}\,,\qquad (T-T_c)\ll T_c\,.
\eqlabel{fcrit}
\end{equation}

\subsection{Quasinormal sound mode}
Sound quasinormal modes in the $\caln=2^*$ black brane geometry involve coupled fluctuations of gauge-invariant 
metric fluctuations $Z_H$ and the gauge-invariant fluctuations of two scalar fields $\{Z_\a,Z_\chi\}$, see \cite{bbs} for details.
The spectrum of quasinormal modes is determined \cite{ks} by imposing on $\{Z_H,Z_\a,Z_\c\}$ 
an incoming wave boundary condition at the horizon, and requiring vanishing of the non-normalizable modes for $\{Z_H,Z_\a,Z_\c\}$ 
near the boundary. In the hydrodynamic limit $\ww\to 0$, $\qq\to 0$ with $\frac \ww\qq$ kept fixed, this leads to the following 
perturbative expansions
\begin{equation}
\begin{split}
Z_H=&(1-x)^{-i\ww}\biggl(z_{H,0}+i\qq\ z_{H,1}+\calo(\qq^2)\biggr)\,,\\
Z_\a=&(1-x)^{-i\ww}\biggl(z_{\a,0}+i\qq\ z_{\a,1}+\calo(\qq^2)\biggr)\,,\\
Z_\c=&(1-x)^{-i\ww}\biggl(z_{\c,0}+i\qq\ z_{\c,1}+\calo(\qq^2)\biggr)\,,
\end{split}
\eqlabel{expan}
\end{equation}
with the following boundary conditions on $\{z_{H,i},z_{\a,i},z_{\c,i}\}$:    
\begin{equation}
\begin{split}
&\lim_{x\to 1_-} z_{H,0}=1\,,\qquad \lim_{x\to 1_-}z_{H,1}=0\,,\qquad \lim_{x\to 1_-} z_{\a,i}=\lim_{x\to 1_-} z_{\c,i}={\rm finite}\,,\\
&z_{H,i}=\calo(x)\,,\qquad z_{\a,i}= \calo\left(x^{1/2}\right)\,,\qquad z_{\c,i}=\calo\left(x^{3/4}\right)\,,
\qquad {\rm as}\qquad x\to 0_+\,.
\end{split}
\end{equation}
Additionally, we parameterize dispersion of the lowest quasinormal mode as  
\begin{equation}
\ww=\frac{1}{\sqrt{3}}\ \qq\ \b_1-\frac i3\ \qq^2\ \b_2+\calo(\qq^3)\,,
\eqlabel{disps}
\end{equation}
where $\b_i=\beta_i(\mu,\nu)$. In the conformal case, $\mu\to 0$ (with $\nu=0$) we expect \cite{ss}
\begin{equation}
\lim_{\mu\to 0}\b_i(\mu,0)=1\,.
\eqlabel{cftl}
\end{equation}
 
Identifying \eqref{sound} with \eqref{disps} and using the universal result  for the shear viscosity \eqref{etas} we find
\begin{equation}
\left(\frac 13-c_s^2\right)=\frac 13\ (1-\b_1^2)\,,\qquad \frac \zeta\eta=\frac 43\ (\b_2-1)\,.
\eqlabel{betai}
\end{equation}

\subsubsection{Leading order in the hydrodynamic approximation}
To leading order in the hydrodynamic approximation, wave functions of the gauge-invariant fluctuations $\{z_{H,0},z_{\a,0},z_{\c,0}\}$ 
satisfy the following equations
\begin{equation}
\begin{split}
0=&z_{H,0}''+\calc_{101}\ z_{H,0}'+\calc_{102}\ z_{H,0}+\calc_{103}\ z_{\a,0}+\calc_{104}\ z_{\c,0}\,,\\
0=&z_{\a,0}''+\calc_{201}\ z_{\a,0}'+\calc_{202}\ z_{H,0}'+\calc_{203}\ z_{H,0}+\calc_{204}\ z_{\a,0}+\calc_{205}\ z_{\c,0}\,,\\
0=&z_{\c,0}''+\calc_{301}\ z_{\c,0}'+\calc_{302}\ z_{H,0}'+\calc_{303}\ z_{H,0}+\calc_{304}\ z_{\a,0}+\calc_{305}\ z_{\c,0}\,,
\end{split}
\eqlabel{order0}
\end{equation}
where connection coefficients $\calc_{i0j}$ are nonlinear functionals of $\{\r_6,c,a\}$ with explicit dependence on $x$
and $\b_{12}\equiv \b_1^2$:
\begin{equation}
\calc_{i0j}=\calc_{i0j}\biggl[\{\r_6,c,a\};\ x;\ \b_{12}\biggr]\,.
\eqlabel{calc0}
\end{equation}
Using \eqref{asbound}, \eqref{ashor}, for each set $\cals_{background}$ we construct the asymptotic solution of 
\eqref{order0}. Near the boundary, $x\to 0_+$ we find
\begin{equation}
\begin{split}
&z_{H,0}=x\ z_{H,2,0}^{(0)}+\calo\left(x^2\ln x\right)\,,\\
&z_{\a,0}=x^{1/2}\ z_{\a,1,0}^{(0)}+\calo\left(x\ln x\right)\,,\qquad z_{\c,0}=x^{1/4}\biggl(x^{1/2}\ z_{\c,1,0}^{(0)}
+\calo\left(x\ln x\right)\biggr)\,,
\end{split}
\eqlabel{sound1}
\end{equation}
and near the horizon, $y\equiv 1-x\to 0_+$
\begin{equation}
\begin{split}
z_{H,0}=1+\calo(y^2)\,,\qquad z_{\a,0}=q_0^{(0)}+\calo(y^2)\,,\qquad z_{\c,0}=x_0^{(0)}+\calo(y^2)\,.
\end{split}
\eqlabel{sound2}
\end{equation}
Thus, altogether we have six new parameters:
\begin{equation}
\cals_{background}\qquad \Longrightarrow\qquad \biggl\{\b_{12},z_{H,2,0}^{(0)},z_{\a,1,0}^{(0)},z_{\c,1,0}^{(0)},q_0^{(0)},x_0^{(0)}\biggr\}\,,
\eqlabel{parsou}
\end{equation}
precisely what is necessary to construct a unique solution for 
$\{z_{H,0},z_{\a,0},z_{\c,0}\}$ for a given $\cals_{background}$.

We use numerical techniques developed in \cite{abk} to general data sets
\begin{equation}
\cals_{sound}\equiv \biggl\{\cals_{background};\ \b_{12},z_{H,2,0}^{(0)},z_{\a,1,0}^{(0)},z_{\c,1,0}^{(0)},q_0^{(0)},x_0^{(0)}\biggr\}\,.
\eqlabel{defss}
\end{equation}

The data sets $\cals_{sound}$ are available from the authors upon request.

\subsubsection{The first subleading order in the hydrodynamic approximation}
To the first subleading order in the hydrodynamic approximation, 
wave functions of the gauge-invariant fluctuations $\{z_{H,1},z_{\a,1},z_{\c,1}\}$ 
satisfy the following equations
\begin{equation}
\begin{split}
0=&z_{H,1}''+\calc_{111}\ z_{H,1}'+\calc_{112}\ z_{H,1}+\calc_{113}\ z_{\a,1}+\calc_{114}\ z_{\c,1}+\calc_{115}\ z_{H,0}'+\calc_{116}\ z_{H,0}
\\&+\calc_{117}\ z_{\a,0}+\calc_{118}\ z_{\c,0}\,,\\
0=&z_{\a,1}''+\calc_{211}\ z_{\a,1}'+\calc_{212}\ z_{H,1}'+\calc_{213}\ z_{H,1}+\calc_{214}\ z_{\a,1}+\calc_{215}\ z_{\c,1}+\calc_{216}\ z_{\a,0}'
+\calc_{217}\ z_{H,0}'\\
&+\calc_{218}\ z_{H,0}+\calc_{219}\ z_{\a,0}+\calc_{2110}\ z_{\c,0}\,,\\
0=&z_{\c,1}''+\calc_{311}\ z_{\c,1}'+\calc_{312}\ z_{H,1}'+\calc_{313}\ z_{H,1}+\calc_{314}\ z_{\a,1}+\calc_{315}\ z_{\c,1}+\calc_{316}\ z_{\c,0}'
+\calc_{317}\ z_{H,0}'\\
&+\calc_{318}\ z_{H,0}+\calc_{319}\ z_{\a,0}+\calc_{3110}\ z_{\c,0}\,,
\end{split}
\eqlabel{order1}
\end{equation}
where connection coefficients $\calc_{i1j}$ are nonlinear functionals of $\{\r_6,c,a\}$ with explicit dependence on $x$
and $\{\b_{12},\b_2\}$:
\begin{equation}
\calc_{i1j}=\calc_{i1j}\biggl[\{\r_6,c,a\};\ x;\ \{\b_{12},\b_2\}\biggr]\,.
\eqlabel{calc1}
\end{equation}

Using \eqref{asbound}, \eqref{ashor}, \eqref{sound1}, \eqref{sound2}, for each set $\cals_{sound}$ we construct the asymptotic solution of 
\eqref{order1}. Near the boundary, $x\to 0_+$ we find
\begin{equation}
\begin{split}
&z_{H,1}=x\ z_{H,2,0}^{(1)}+\calo\left(x^2\ln x\right)\,,\\
&z_{\a,1}=x^{1/2}\ z_{\a,1,0}^{(1)}+\calo\left(x\ln x\right)\,,\qquad z_{\c,1}=x^{1/4}\biggl(x^{1/2}\ z_{\c,1,0}^{(1)}
+\calo\left(x\ln x\right)\biggr)\,,
\end{split}
\eqlabel{att1}
\end{equation}
and near the horizon, $y\equiv 1-x\to 0_+$
\begin{equation}
\begin{split}
z_{H,1}=0+\calo(y^2)\,,\qquad z_{\a,1}=q_0^{(1)}+\calo(y^2)\,,\qquad z_{\c,1}=x_0^{(1)}+\calo(y^2)\,.
\end{split}
\eqlabel{att2}
\end{equation}
Thus, altogether we have six new parameters:
\begin{equation}
\cals_{sound}\qquad \Longrightarrow\qquad \biggl\{\b_{2},z_{H,2,0}^{(1)},z_{\a,1,0}^{(1)},z_{\c,1,0}^{(1)},q_0^{(1)},x_0^{(1)}\biggr\}\,,
\eqlabel{paratt}
\end{equation}
precisely what is necessary to construct a unique solution for 
$\{z_{H,1},z_{\a,1},z_{\c,1}\}$ for a given $\cals_{sound}$.

We use numerical techniques developed in \cite{abk} to general data sets
\begin{equation}
\cals_{attenuation}\equiv \biggl\{\cals_{sound};\ \b_{2},z_{H,2,0}^{(1)},z_{\a,1,0}^{(1)},z_{\c,1,0}^{(1)},q_0^{(1)},x_0^{(1)}\biggr\}\,.
\eqlabel{defsa}
\end{equation}

The data sets $\cals_{attenuation}$ are available from the authors upon request.

\section{Bulk viscosity of $\caln=2^*$ plasma}
Given $\cals_{attenuation}$ we have all the necessary data to study bulk viscosity of strongly coupled $\caln=2^*$ plasma. 
Of primary interest to us are the bulk viscosity bound conjecture \cite{b1}, and the behaviour of bulk viscosity near the phase 
transition.   

\subsection{Bulk viscosity bound}

\begin{figure}[t]
 \hspace*{-20pt}
\begin{center}
\psfrag{x}{\raisebox{-1ex}{\hspace{-0.3cm}$\left(\frac 13-c_s^2\right)$}}
\psfrag{ze}{\raisebox{0ex}{\hspace{0.0cm}$\frac{\zeta}{\eta}$}}
 \includegraphics[width=6.0in]{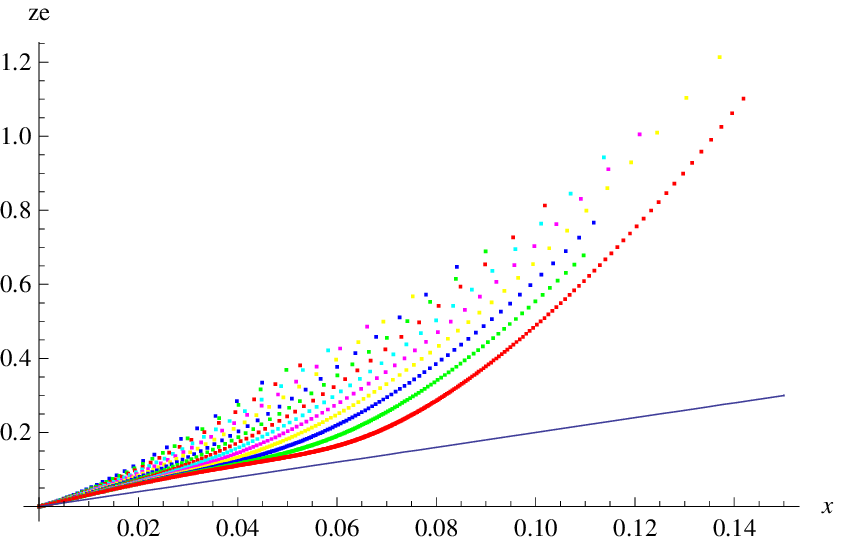}
\caption{Ratio of viscosities  $\frac{\zeta}{\eta}$ versus the speed of sound in 
$\caln=2^*$ gauge theory plasma with 
mass deformation parameter $\nu\equiv \frac{m_f^2}{m_b^2}\in[0.2,0.9]$, with intervals $\Delta\nu=0.05$. 
The solid line represents the bulk viscosity bound \eqref{bound}.}
\end{center}
\label{fig1}
\end{figure}

\begin{figure}[t]
 \hspace*{-20pt}
\begin{center}
\psfrag{x}{\raisebox{-1ex}{\hspace{-0.3cm}$\left(\frac 13-c_s^2\right)$}}
\psfrag{ze}{\raisebox{0ex}{\hspace{0.0cm}$\frac{\zeta}{\eta}$}}
 \includegraphics[width=6.0in]{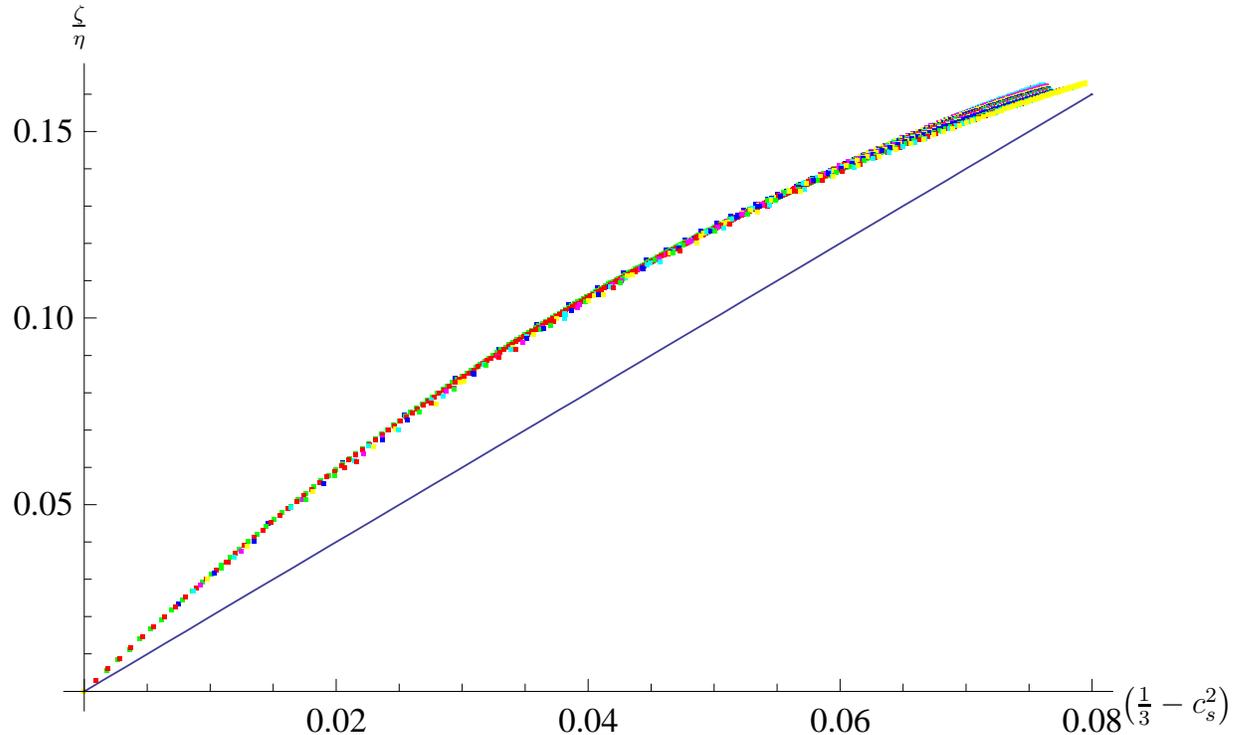}
\caption{Ratio of viscosities  $\frac{\zeta}{\eta}$ versus the speed of sound in 
$\caln=2^*$ gauge theory plasma with 
mass deformation parameter $\nu\equiv\frac{m_f^2}{m_b^2}\in\{1,1.05,[1.1,2.1]_{\Delta \nu=0.1},[3,6]_{\Delta \nu=1}\}$. 
The solid line represents the bulk viscosity bound \eqref{bound}.}
\end{center}
\label{fig2}
\end{figure}

Fig.~1 represents the ratio  of bulk to shear viscosities $\zeta/\eta$ as a function of $(\ft 13-c_s^2)$ 
for $\caln=2^*$ plasma with mass deformation parameter $\nu$ ranging from $\nu=0.2$ to $\nu=0.9$ with 
intervals of $\Delta\nu=0.05$. Different color sets of points represent different values of $\nu$.
The solid blue line represents the bulk viscosity bound \eqref{bound}. We verified that in the vicinity 
of $c_s^2=\ft 13$ (which corresponds to a high-temperature regime of $\caln=2^*$ plasma), the results are 
in excellent agreement with \eqref{n2r1}, \eqref{n2r2}.
Although in this paper we truncated the plots to a near-conformal regime\footnote{The rapid growth of bulk 
viscosity indicates that if any violation of the bound \eqref{bound} would occur, it would occur in a near-conformal 
regime.}, available data sets $\cals_{attenuation}$ allow us to study the  behaviour of bulk viscosity 
near the critical point $T_c(\nu)$. Much like for the $\nu=0$ case discussed in \cite{b1}, for each value of $\nu<1$
we observe a rapid, power-law like, growth of $\zeta/\eta$ in the vicinity of the critical point.
Nonetheless, this ratio is finite precisely at $T=T_c$ (see Fig.~3 below). 
The results are qualitatively identical to the $\nu=0$ case discussed in \cite{b1}.  

Fig.~2 represents the ratio  of bulk and shear viscosities $\zeta/\eta$ as a function of $(\ft 13-c_s^2)$ 
for $\caln=2^*$ plasma with mass deformation parameter $\nu=\{1,1.05\}$, also in the range from
$\nu=1.1$ to $\nu=2.1$ with intervals of $\Delta\nu=0.1$, and in the range from
$\nu=3$ to $\nu=6$ with intervals of $\Delta\nu=1$. Different color sets of points represent different values of $\nu$.
The solid blue line represents the bulk viscosity bound \eqref{bound}. Again, we verified that in the vicinity 
of $c_s^2=\ft 13$ (which corresponds to a high-temperature regime of $\caln=2^*$ plasma), the results are 
in excellent agreement with \eqref{n2r1}, \eqref{n2r2}. Notice the accumulation of the points as data sets 
approach the bound. Much like for the $\nu=1$ case discussed in \cite{b1}, this regime corresponds to  a low-temperature 
$\caln=2^*$ plasma regime.  Specifically, there we have $m_b/T\sim 5\cdots 10$. Much like in \cite{b1} we can extrapolate 
the speed of sound and the viscosity ratio to $T\to 0$. We find that the endpoints of such extrapolations land on 
(or slightly above) the bulk viscosity bound line. 

It is very interesting to explore the thermodynamics of $\caln=2^*$
plasma (for $\nu>1$) at extremely low temperature\footnote{Our current numerical algorithms become 
very slow as low temperatures.}, ideally, as $T\to 0$. So far we have no indication that such a zero temperature limit 
would be singular. On the other hand, we can not exclude the presence of some 
exotic  instabilities/phase transitions at very low temperatures, see \cite{bl} for more details.

\subsection{Bulk viscosity at a critical point}

\begin{figure}[t]
 \hspace*{-20pt}
\begin{center}
\psfrag{nu}{\raisebox{-1ex}{\hspace{-0.3cm}$\nu\equiv \ft {m_f^2}{m_b^2}$}}
\psfrag{ze}{\raisebox{0ex}{\hspace{0.0cm}$\frac{\zeta}{\eta}\bigg|_{critical}$}}
 \includegraphics[width=6.0in]{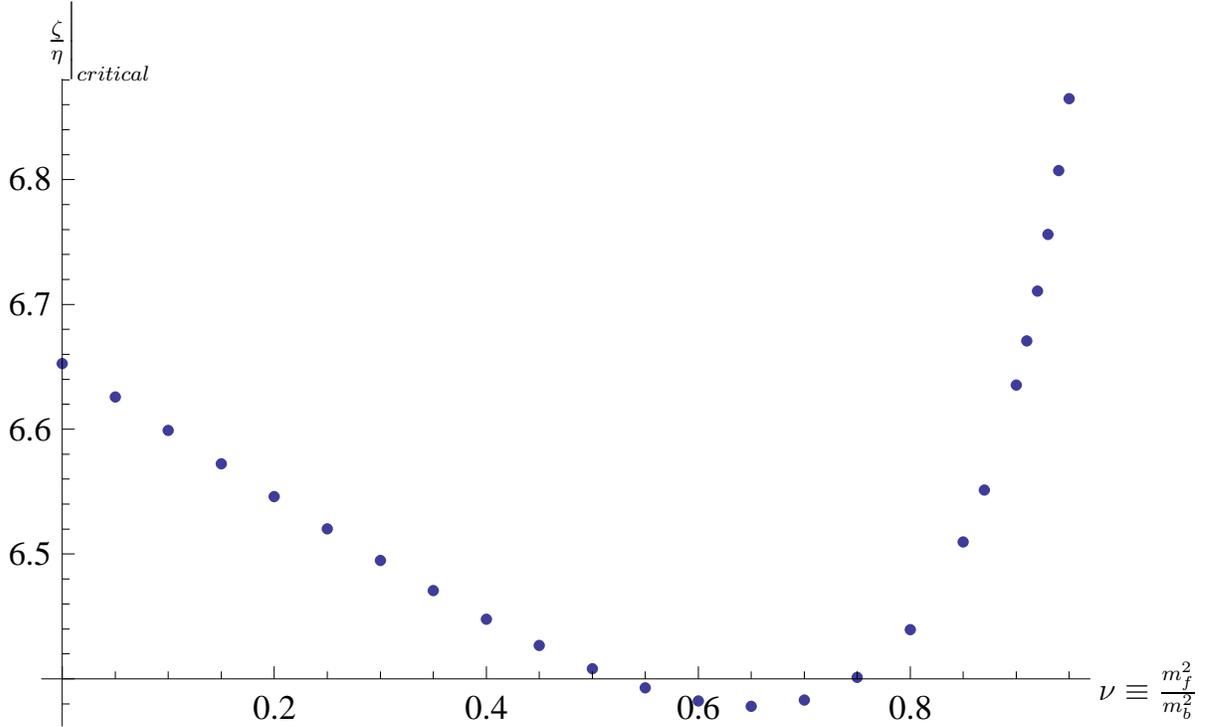}
\caption{Ratio of viscosities  $\frac{\zeta}{\eta}$ at $T=T_c$ versus the 
mass deformation parameter $\nu$ of $\caln=2^*$ plasma.}
\end{center}
\label{fig3}
\end{figure}

We already mentioned that $\caln=2^*$ plasma with mass deformation parameter $0\le \nu<1$ undergoes a phase transition, 
which appears to be in the universality class of the mean-field tricritical point. At such a phase transition, specific 
heat diverges as $c_V\sim |1-T_c/T|^{-1/2}$. Fig.~3 represents the ratio of bulk to shear viscosities 
$\zeta/\eta$ as a function of $\nu\equiv \ft {m_f^2}{m_b^2}$ at the critical temperature $T=T_c(\nu)$. 
We took the $\nu=0$ result from \cite{b1}. Notice the rapid growth of $\frac{\zeta}{\eta}\bigg|_{critical}$
as $\nu\to 1_-$. Actually, such a behavior is expected if $\nu=1$ is the end point at which  transitions cease to occur,
and for $\nu \ge 1$ the ratio of $\zeta/\eta$ is relatively close to the bound \eqref{bound}  
down to rather low temperatures.

\section{Conclusion}
In this paper we study the bulk viscosity of $\caln=2^*$ gauge theory plasma at strong coupling as a function of 
temperature and for various masses, $\nu\equiv m_f^2/m_b^2$. 
In all cases, we find that the viscosity bound \eqref{bound} is satisfied. 
Similar to a thermal state of $\caln=2^*$ plasma with $\nu=0$, we find that, while the bulk viscosity of the $\caln=2^*$
plasma  grows rapidly in the vicinity of the critical point for $0<\nu<1$, it is finite precisely at $T=T_c(\nu)$.

We did not discuss in this paper gauge theory/string theory duality-motivated phenomenological approaches to transport properties
(see however \cite{gubser}). We also did not discuss potential applications to RHIC/LHC physics (see however \cite{kapusta}).

In the future it would be interesting to study transport properties of the Klebanov-Strassler \cite{ks2} cascading 
plasma. The preliminary work necessary for such analysis already appeared in the literature \cite{ks1,abk,ks3}.

\section*{Acknowledgments}
We would like to thank Krishna Rajagopal for valuable discussions. Research at Perimeter Institute is
supported by the Government of Canada through Industry Canada and by
the Province of Ontario through the Ministry of Research \&
Innovation. AB gratefully acknowledges further support by an NSERC
Discovery grant and support through the Early Researcher Award
program by the Province of Ontario. CP acknowledges support by NSERC.


\begin{thebibliography}{99}
\bibitem{mistake} There is a mistake in Eq.~(4.37) in \cite{bbs}: correspondingly to the connection 
coefficient of $dZ^0_\psi/dx$ in Eq.~(4.35), 
the connection coefficient of $dZ^1_\psi/dx$ in Eq.~(4.37) must  be $12 x^2(x^2-1)^2$. Fixing this mistake 
leads to the value of $\beta_f^\Gamma$ presented.
\end{thebibliography}
\end{document}